# Understanding the role of single-board computers in engineering and computer science education: A systematic literature review

**Jonathan Álvarez Ariza[1], Heyson Baez[2]**
[1,2]Department of Technology in Electronics, Engineering Faculty, Corporación Universitaria Minuto de Dios-UNIMINUTO, 111021, Bogotá, Colombia
[1]**Email:** jalvarez@uniminuto.edu
[2]**Email:** heysonbaez@gmail.com

## Abstract

In the last decade, Single-Board Computers (SBCs) have been employed more frequently in engineering and computer science both to technical and educational levels. Several factors such as the versatility, the low-cost, and the possibility to enhance the learning process through technology have contributed to the educators and students usually employ these devices. However, the implications, possibilities, and constraints of these devices in engineering and Computer Science (CS) education have not been explored in detail. In this systematic literature review, we explore how the SBCs are employed in engineering and computer science and what educational results are derived from their usage in the period 2010-2020 at tertiary education. For that, 154 studies were selected out of $n=605$ collected from the academic databases Ei Compendex, ERIC, and Inspec. The analysis was carried-out in two phases, identifying, e.g., areas of application, learning outcomes, and students and researchers' perceptions. The results mainly indicate the following aspects: (1) The areas of laboratories and e-learning, computing education, robotics, Internet of Things (IoT), and persons with disabilities gather the studies in the review. (2) Researchers highlight the importance of the SBCs to transform the curricula in engineering and CS for the students to learn complex topics through experimentation in hands-on activities. (3) The typical cognitive learning outcomes reported by the authors are the improvement of the students' grades and the technical skills regarding the topics in the courses. Concerning the affective learning outcomes, the increase of interest, motivation, and engagement are commonly reported by the authors.

## 1. Introduction

Since the first model of Raspberry Pi was released in 2012 [1], the interest and proliferation of the Single-Board Computers (SBCs) in engineering and computer science have increased over the years. This trend has been supported by factors such as the reduction in the manufacturing costs, the evolving of the electronic components, and an imperative need to reform the traditional curricula in engineering and computer science programs, making them more interesting, inclusive, and accessible according to the learning needs of the students. From a technical perspective, a SBC is a small computer whose main components, e.g., processor and memory are integrated in a single System on Chip (SoC), allowing the students to interact with hardware elements such as sensors or actuators in applications that encompass computing, robotics, Internet of Things (IoT), among others [2–4]. This interaction is known as physical computing [5]. While the concept of SBC is not rather new, it was born in 1976 with the Dyna-micro computer [3], its utilization in engineering and computer science education is recent with most part of the proposed investigations situated in the last decade. As it is well-documented in the literature, the attractiveness of these computers lies in several technical features such as versatility, low-cost, low power consumption, multipurpose and an extensive open-source community of developers and makers that help to maintain their applications and functionalities [6]. Moreover, the proliferation of the Do It Yourself (DIY) movement that has the philosophy to create, modify and repair certain technologies without the explicit assistance of professionals [7,8] has led to the popularization of the SBCs in higher education.

However, despite these important features, the simple incorporation of technologies, e.g., SBCs in the classrooms neither guarantees learning nor creates educational spaces in which the students can interact and collaborate. In part, as we discovered in this review, there seems to be a technocentric view of technology in a large part of the analyzed articles, even in those better cited. Educators may be giving a central relevance to the technical aspects entailed to the SBCs rather than observing the outcomes and effects that they produce in engineering and computer science education. As Seymour Papert [9, p. 23] observes,

> The context for human development is always a culture, never an isolated technology. In the presence of computers, cultures might change and with them people's ways of learning and thinking. But if you want to understand (or influence) the change, you have to center your attention on the culture– not on the computer.

Although the incorporation of SBCs in the curricula of engineering and Computer Science (CS) continues increasing and despite the plethora of technical reports about them, there exists a current gap of literature and critical reviews that synthesize how these computers are being used and what implications they have in higher education. There are few studies that have tried to tackle this issue, for instance, the studies in [10–13] synthetize several implications of SBCs in the educational arena. In particular, the systematic mapping review in [10] explores some SBCs such as Raspberry Pi and BeagleBone in the educational context. The study provides methodological guidance to address the vast number of proposals which in their majority are oriented towards new approaches to solve a problem in education as well as to improve the teaching methods and student engagement.

Thus, in order to elicit how the SBCs are used in engineering and CS education at the tertiary levels (undergraduate, master, and PhD degrees), we performed a Systematic Literature Review (SLR) according to the process indicated by [14] based on $n=605$ papers selected from the databases Ei Compendex, Inspec, and Education Resources Information Center (ERIC) in the period 2010-2020. This timeframe matches with the rise of the SBCs in higher education. The SLR was divided into two stages. In the first one, we described an overall analysis to identify top-cited articles, the distributions of the proposals in terms of publication years and tertiary levels, and the current educational areas in which the SBCs are utilized. In the second one, we identified through the empirical technique of content analysis [15,16], the educational methodologies, the learning outcomes, and both researchers and students' perceptions about the implications of the SBCs in the educational process.

Given the previous elements, the main contributions of this paper are focused on the following aspects: (1) Providing a state of art about the current educational areas in which SBCs are being employed. (2) Informing researchers and practitioners about the educational methodologies and learning outcomes that have been achieved through the usage of SBCs. (3) Describing the point of view of researchers and students about the SBCs and their implications for learning, motivation, and engagement, and (4) suggest future directions for researchers based on the synthesis of the literature in the field of the SBCs in engineering and CS education.

The remainder of the paper is structured as follows. Section 2 describes the background focusing on the concept of Physical Computing. Section 3 explains the followed method and tools employed for the SLR. The synthesis and outcomes to answer the research questions in the SLR are explained in section 4. Finally, section 5 and 6 outline the limitations and the conclusions of this review, respectively.

## 2. Background and motivation of this study

Physical Computing (PhyC) is a key concept in the SLR because it has allowed that SBCs will be used in higher education, contributing to change the traditional way to make computing. Essentially, computing is not an isolated field, it requires in many cases the interaction of the designed algorithms with hardware devices to get information about the processes or systems that are handled in engineering or CS. In the educational level, PhyC has been incorporated into the education arena during the last two decades to promote and enhance the learning process of the students specially regarding programming, problem solving and computational thinking [17–19]. The new generation of students in engineering and CS need to engage with hands-on activities that foster their creativity and collaboration not only in the traditional screen-based way to make programming but also employing both physical artifacts and technologies [20]. Nevertheless, the definition of this concept is blurred given the multiple approaches to it. PhyC is defined in terms of tangible interfaces [21], cyber-physical systems [22], or pervasive computing [23]. While it is true that PhyC is a transversal concept to these fields, the scope of them is broader in computing education. Hodges et al. [20] define PhyC as a combination of hardware and software to build physical systems that sense and interact with the real world. Greenwold [24] states that PhyC is a type of human interaction with machines in which they manipulate real objects and spaces. Øritsland and Buur [25] indicate that PhyC is an interaction where the term interaction is understood as "an iterative process of listening, thinking, and, speaking between two or more actors". In those definitions at least three elements are unveiled. Firstly, PhyC needs interaction with computing and artifacts such as sensors or actuators in the real world. This interaction is mainly created by transductors which are elements that transform the sensed variables to an electric equivalent in voltage or current. Secondly, this interaction or mediation from a constructivist point of view benefits the cognitive, perceptual, and social skills, in this case of the students. Thirdly, PhyC is essentially an activity with computers to manipulate objects in the real world. PhyC is closely related to SBCs because one of the main features of these devices is their possibility to interact with the environment through General Purpose I/O (GPIOs) and sensors, processing the information according to the design purposes and requirements. Complementary, in terms of the author Verenikina [26, p. 21] that explores the Vygotskian constructivism and Technology Enhanced Learning (TEL) "a physical tool can be seen as an instrument of labour, a thing which is interposed between a person and the object of their labour and which serves as the conductor of their activity". For this case, the author refers that the object for the students in the educational context is to promote and enhance their learning.

Besides, the educational tenets of the PhyC are mainly based on the notion of constructivism and bricolage. As described, educators have felt the need to make the curricula of engineering and CS more inclusive and interesting for the students. In this scenario, the dominant theory of learning is constructivism which claims that knowledge is constructed actively by the students [27]. Under the umbrella of constructivism, aspects such as experimentation, learning by doing, and bricolage take more relevance in the context of engineering and CS. Regarding bricolage, it is a term created by the anthropologist Lévi-Strauss to name the science of the concrete in primitive societies in counterpart to the abstract European science [27]. The term is rescued and reconstructed by Seymour Papert in his theory of learning known as constructionism which gathers the foundations of constructivism and experimentation through hands-on activities made by the students in the field of programming and computer interaction [28]. Constructionism deals with the idea that knowledge and learning should be created by the students instead of being a simple act of information transmission between teachers and learners. Students learn better when they can construct tangible things, and they can see the effect of their algorithms or programs to create generalizations and abstractions [29]. PhyC has had special attention, e.g., in the field of robotics because students can make a transition between black-box devices towards white-box designs in which they can construct their robots and algorithms to control them [29,30]. In this regard, there have been several initiatives for children's education such as the revisited version of LOGO with Python language denominated *Phogo* [31] and the low-cost DIY platform *PiBot* conceived for STEM education [32]. Likewise, PhyC has demonstrated to be a mechanism to reduce the gender gap regarding confidence in programming abilities in courses of introduction to programming [33]. Women learn better programming, for example, in educational environments that use robots [34]. Between the benefits to include PhyC in the classrooms are highlighted aspects such as increasing motivation and self-confidence, fostering creativity, collaboration, inclusion, learning by doing, and engagement [20].

Nonetheless, although all these aspects suggest positive outcomes for learning in the students, several researchers have reported problems about the application of bricolage in education from the perspectives of students and educators. For instance, Ben-Ari [27] describes that the trend "*try and see what happens*" can be beneficial for novice programmers but could be risky for professional programming where planning in software engineering must be learned and practiced. Hatton [35] argues that bricolage is a response to contingencies that neither seek new theoretical structures nor a better grasping or technical competence. Indeed, bricolage could have negative impacts on the labor of educators. When educators are exposed to constraints, conservatism, a lack of creativity or even improvisation can appear in the classroom.

Thus, to clarify the previous points that in part seem contradictory, our interest with the SLR is to provide a perspective of application regarding the SBCs in engineering and CS education based on the evidence. In this sense, we search to answer the overall question *How are the SBCs employed in engineering and CS education?* With the information of the SLR, we hope that educators and researchers create future studies and methodologies that help to comprehend the implications of the SBCs employment for both students and educators in higher education.

## 3. Method

Systematic reviews search to answer a set of questions to identify and reveal current gaps, to contrast hypotheses, or to expand the scope of topics in a particular knowledge area [14]. The information provided by the systematic reviews allows stakeholders, practitioners, and researchers to take decisions and plan future studies to close breaches based on the collected evidence [36]. In this way, the SLR was carried out with the steps indicated by Gough et al. [14], which are: (1) Pose the research questions and the methodology; (2) Define the search and the screening according to eligibility criteria; (3) Code to match or build a conceptual framework; (4) Apply quality appraisal criteria; (5) Synthesize of the review; and (6) Interpret and communicate the findings.

As mentioned, the main purpose of the review is to identify how the SBCs are employed in engineering and CS education and what learning outcomes are supported by their use. To address this aim, we formulated the following research questions that oriented our work:

- **RQ1.** What are the distributions of the proposals in terms of publication years, top-cited articles, and tertiary levels?
- **RQ2.** What are the educational areas of engineering and computer science in which the SBCs are used?
- **RQ3.** What are the main features of the SBCs that support learning and teaching in engineering and computer science education?
- **RQ4.** What are the learning outcomes and main findings reported by the authors which are supported by the usage of SBCs?

The SLR was conducted by the PRISMA (Preferred Reporting Items for Systematic Reviews and Meta-Analyses) guidelines [37] which provide order in the phases of identification, screening, eligibility, and inclusion. Figure 1 illustrates the PRISMA flow diagram for the review.

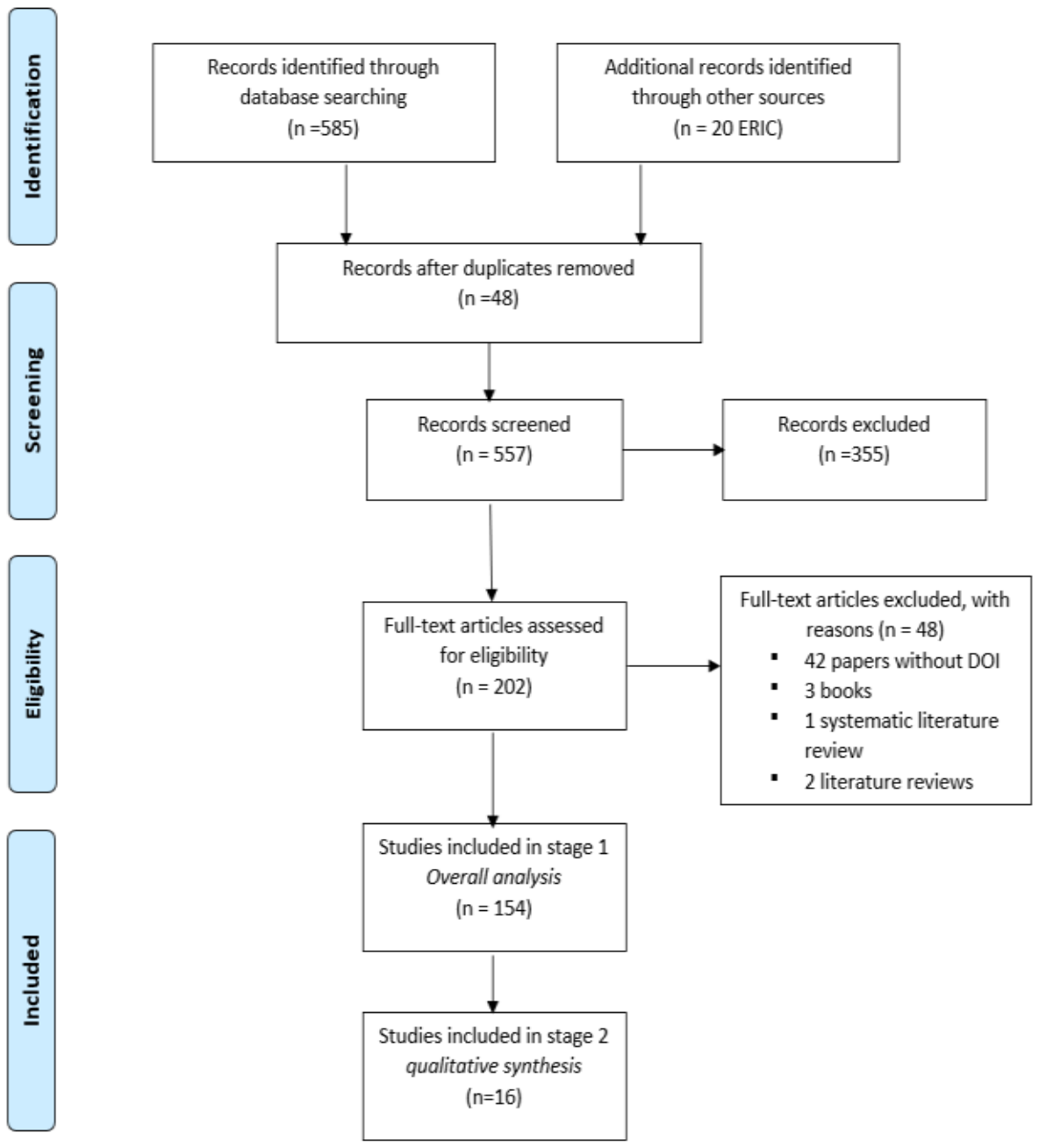

**Figure 1**. PRISMA flow diagram for the study adapted from reference [37].

### *3.1 Searching Criteria*

According to the research questions, the search was bounded to the interval 2010-2020 that corresponds to a timeframe in which the major part of the SBCs started to be used in the educational arena. 585 Studies were retrieved with the tool Engineering Village from Elsevier [38]. This tool was employed because it gathers several relevant academic databases in the Engineering field such as Ei Compendex and Inspec, which helped us to centralize and organize the information. Additionally, a search was made in the database Education Resources Information Center (ERIC), identifying 20 additional records.

For the search, several attempts to refine the searching strings were made until the final version. In the attempts, we modified the Boolean connectors (AND, OR, NOT), observing the response of the tool Engineering Village and ERIC with the possible overlapped terms. For instance, the terms of machine learning and deep learning were concurrent in our search. When we examined these papers, all of them were non-educational proposals regarding the scope of the review. So, we excluded these terms from the search string. A similar rationale was performed to exclude proposals outside of higher education and the educational framework. With these elements, the final searching strings are consigned in Table 1.

With the criteria depicted in Table 1, the searching process was made by two researchers (A, B) independently to contrast the results and to avoid possible bias. Information retrieved with the final search string was downloaded in Research Information Systems (.RIS) citation files and organized in Mendeley to eliminate duplicates.

**Table 1.** Description of searching criteria for the SLR.

| ***Aspect*** | ***Description*** |
|---|---|
| **Date of search** | March 18, 2020 |
| **Timeframe** | 2010-2020 |
| **Databases; Searching tools** | ERIC, Ei Compendex, Inspec; Elsevier Engineering Village |
| **Searching String** | **Engineering Village (Ei Compendex, Inspec):** *("raspberry pi" OR "beaglebone" OR "beagleboard" OR "odroid" OR "intel edison" OR "orange pi" OR "tinker board" OR "intel galileo") AND (education OR teaching OR learning NOT "machine learning" NOT "deep learning" NOT "high school" NOT "primary school" NOT "elementary school" NOT "secondary school")*<br><br>**ERIC:** *(“raspberry pi” OR “beaglebone” OR “beagleboard” OR “odroid” OR “intel edison” OR “orange pi” OR “tinker board” OR “intel galileo”) AND (engineering OR computer science)* |
| **Inclusion Criteria** | **Single Board Computers (SBCs):** Raspberry Pi, Beaglebone, Beagleboard, Odroid, Intel Edison, Orange Pi, Tinker Board, Intel Galileo.<br><br>**Higher education (tertiary levels):** Undergraduate, master, and PhD degrees.<br><br>**Scope:** Engineering and computer science education.<br><br>**Language:** English.<br><br>**Type of study**: Primary research. |
| **Exclusion Criteria** | Not in engineering or computer science education, not in higher education, not in the timeframe, not application of SBCs, not in English, not primary research. |
| **N° of records obtained** | 605 |

### *3.2 Inter-rater reliability, screening, and eligibility*

To reduce the problems of systematic selection bias, 30 papers were randomly selected from the initial papers group. Then, these papers were classified to be included or excluded by two coders (A, B) according to the criteria of Table 1. Cohen's Kappa coefficient (κ) [39] was calculated using IBM SPSS Statistics v.23 to test the inter-rater reliability. Cohen's Kappa coefficient measures the degree of agreement between raters in which the values of 0.4 to 0.6 are fair, 0.6 to 0.75 good, and over 0.75 excellent [40]. In this case, ( $\kappa$=0.867) which is deemed excellent for the process of inclusion and exclusion. Coders (A, B) made the process of screening where $n$=557 articles were assessed for eligibility after the duplicate papers ($n$=48) were removed. The screening process started with the reading title and abstract, applying the inclusion, exclusion criteria in Table 1. When an ambiguous study was found, the full text was read to clear the different doubts, then, coders (A, B) discussed its inclusion or exclusion. Each Digital Object Identifier (DOI) was tested in both Mendeley and employing the Crossref REST API in Python language [41]. The API allowed to locate the papers with their metadata for the subsequent extraction and analysis. We selected this method to guarantee the availability of all articles. In the eligibility phase, 48 papers were excluded for the reasons indicated in Figure 1. Finally, ($n$=154) studies remained for synthesis.

### *3.3 Data extraction and analysis*

*Stage 1 (Overall Analysis)*: In this stage, we pursue to answer the RQ1-RQ2. For the RQ1, we extracted the metadata corresponding to publication years and cites by article with the Crossref REST API. The API receives the Digital Object Identifier (DOI) of each article of the synthesis group. Afterward, we classified the articles by years and citations. Concerning the tertiary levels, each paper was classified in the levels of undergraduate, master, and PhD degrees. Furthermore, some proposals were classified in the category of persons with disabilities because they can be employed in any of the mentioned tertiary levels. Also, to complement these classifications, the papers were organized into three categories according to the following criteria:

1. Paper describing or proposing a technology with the incorporation of SBCs for educational purposes (TP).
2. Paper reporting lessons learned with the SBCs in the educational context (LL).
3. Paper reporting an empirical study, e.g., a case study with a methodology, assessment, and discussion (CS).

For the RQ2, to guarantee objectivity in the classification of the papers, it was selected the software VOSViewer 1.6.14.0 to analyze the areas in which the SBCs are used in engineering and CS education. VOSViewer is a software for bibliometric analysis based on network data which is focusing on items and clusters with two overall functions: create maps and visualize them [42]. The items are objects of interest, e.g., abstracts, titles, keywords, or authors, and the clusters are patterns that share these items. The analysis with this software was performed with a co-occurrence of minimum 4 keywords in the corpus of the selected papers in the synthesis group. In this process emerged the areas of *laboratories and e-learning, computing education, robotics, Internet of Things (IoT), and persons with disabilities*. For each one of the areas, a state of art was constructed with the most outstanding works.

*Stage 2 (Qualitative Synthesis)*: Searching to answer the questions RQ3-RQ4, we identified through the qualitative technique of content analysis [15,16], the methodologies, learning outcomes, and main educational features of the SBCs presented by the authors. According to [15], content analysis is a research technique that provides an objective description that is systematic and quantitative to interpret the manifest content of the communications. In the content analysis, it must be guaranteed the conditions of completeness, homogeneity, relevance, and exclusivity to make a rigorous examination of the information. Thus, we selected $n=16$ articles with an appropriate evidence of application and assessment that had a quantitative, qualitative, or mixed approaches, all of them case studies. To extract the student opinions, the SBC features, the methodologies, the learning outcomes, and study purposes described by the authors, it was employed the Computer Assisted/Aided Qualitative Data Analysis Software (CAQDAS) *NVivo 12*. In addition, we constructed the Table 6 to complement and summarize the information of this stage. The Table contains the educational methodology, research approach, instruments, participants, study aim, and learning outcomes of each proposal and it is available in the appendix section. As for the learning outcomes, these were classified into learning domains (cognitive and affective) [43]. The cognitive domain is associated with knowledge creation and the development of intellectual abilities and skills. The affective domain is concerned with how students deal with emotions, feelings, values, motivations, or attitudes.

### *3.4 Reporting the results*

Guidelines proposed by Webster and Watson [44] were considered to write the review in aspects such as identifying relevant literature, review based on a concept-centric approach, tables and figures presentation, tone and structure of the synthesis. Results are presented according to the described Research Questions (RQ). Discussion of the review contains a dialog between the findings, educational actors (students and educators), the possibilities and issues to employ the SBCs in Engineering and CS education. Tables 4 to 8, Figure 7, and the link for supplementary materials in the SLR can be found in the appendix section. In the online repository can be found the tables with the description of the studies in each stage, the state of art with the articles selected in the cluster analysis, the *NVivo* coding files, figures, and the scripts employed to extract the information.

## 4. Results and discussion

### *4.1 RQ1. What are the distributions of the proposals in terms of publication years, top-cited articles, and tertiary levels?*

Publications about SBCs in engineering and CS education have increased since 2012 with a relative peak of 35 proposals in 2017 according to Figure 2. This pattern agrees with the development of the Raspberry Pi in its model 3 but also with the appearance and technical evolution of others SBCs on the educational horizon such as BeagleBone, Odroid, Orange Pi, Intel Edison, among others. Also, in this year, authors exposed proposals mainly focused on topics such as IoT, computing education, e-learning, Hardware-In-the-Loop (HIL) simulations, and microcontrollers.

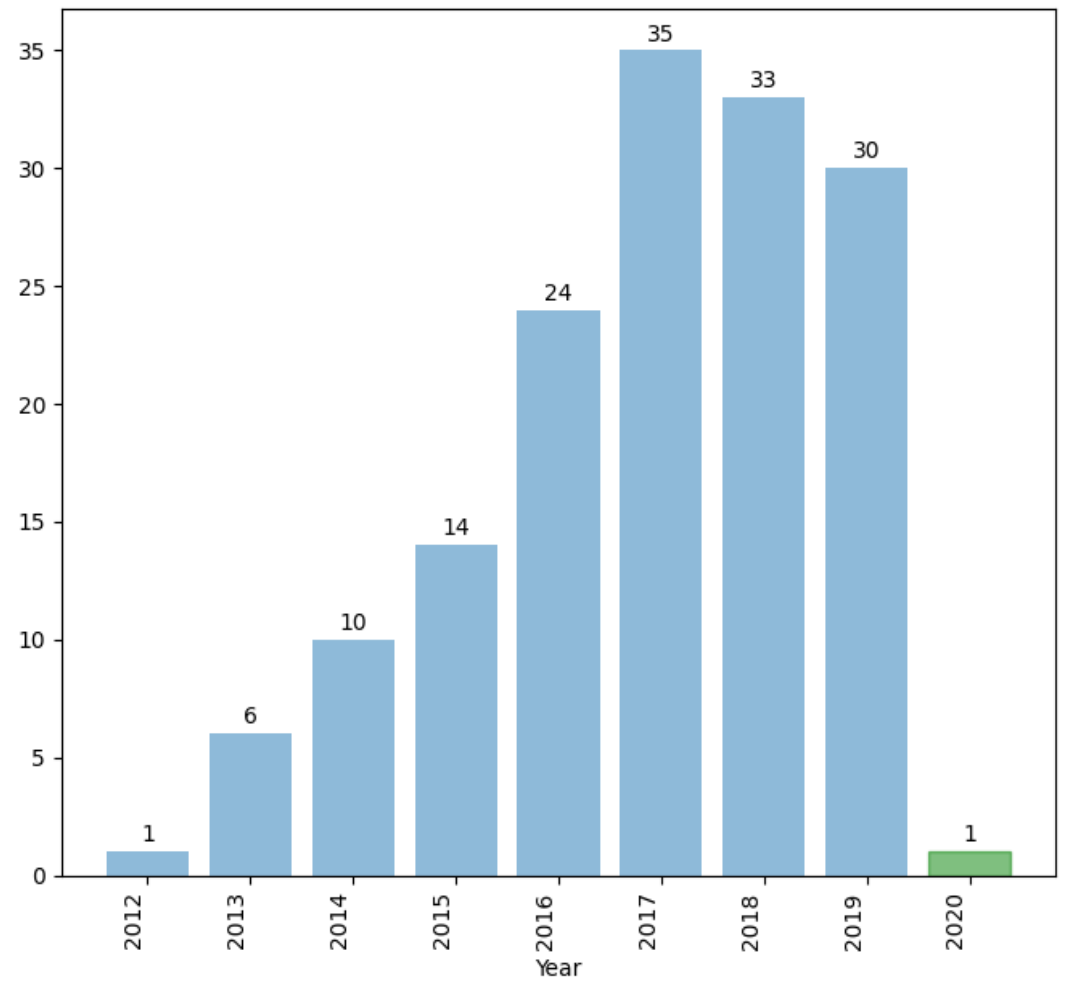

**Figure 2**. Number of publications per year ( *n=154*).

In the timeframe of the review from 154 selected studies, 73.4% came from proceedings or conference articles, 16.9% from journal articles, and 9.7% from book chapters. Table 4 shows the top-cited articles that gather Technological Proposals (TPs), Cases Studies (CSs), and Lessons Learned (LL). In this line, the most cited article describes the development of Iridis-pi, a low-cost cluster for high -performance computing with educational purposes [45]. Besides, in these studies, 112 were TPs (72.72%), 37 CSs (24%), and 5 LL (3.25%).

Table 2 depicts the most cited events or journals with two or more contributions, leading by IFAC-Papers Online with 39 citations. This publication is sponsored by the International Federation of Automatic Control (IFAC) with Elsevier whose main topic is control systems and automation. Publications within the book Advances in intelligent systems and computing from Springer are associated with multidisciplinary topics such as ICT, e-learning, teaching human-centered, human-centric computing, and robotics. This publication occupies second place in cited publications. The 49th proceedings of the Association for Computing Machinery (ACM) in its educational division and the 2017-IEEE Frontiers in Education (FIE) conference occupy third and fourth places in events more cited. At last, the publication International Journal of Online and Biomedical Engineering (iJOE) that cover topics in e-learning in engineering occupies the fifth place.

**Table 2.** Top five of cited events or publications with two or more contributions.

| Event or publication | Publisher | Citations |
|---|---|---|
| 1. IFAC-PapersOnLine | Elsevier | 39 |
| 2. Advances in intelligent systems and computing | Springer | 11 |
| 3. Proceedings of the 49th ACM technical symposium on computer science education | ACM | 9 |
| 4. 2017 IEEE Frontiers in Education (FIE) Conference | IEEE | 7 |
| 5. International Journal of Online Engineering (iJOE) | International Association of Online Engineering | 3 |

Regarding the tertiary levels, 76.62% of the proposals are at the undergraduate, 2.59% at master, and 0.65% shared between PhD and master's degrees. Some proposals were designed for persons with disabilities (3.89%) and they can serve in any of the mentioned tertiary levels. Finally, several proposals (16.23%) did not indicate explicitly their tertiary level.

### *4.2 RQ2. What are the educational areas of engineering and computer science in which the SBCs are used?*

Figure 3 illustrates the map of the clusters produced by VOSViewer based on the papers of the synthesis group (*n*=154). Regarding this map, Figure 4 depicts its density plot. Five overall areas that group the clusters in Figure 3 were identified in the analysis: *Laboratories and e-learning, computing education, robotics, IoT, and persons with disabilities*. The Raspberry Pi was the most concurrent SBC in the analyzed proposals. In addition, Table 5 shows a list of software tools employed by the authors in their studies, and Figure 7 depicts a scatter plot in terms of SBCs, cluster areas, and type of study (TP, CS, or LL). A synthesis with the most outstanding works in each area is illustrated in Table 3. Similarly, a state of art with the description of each study in Table 3 can be found online at the repository in the appendix section.

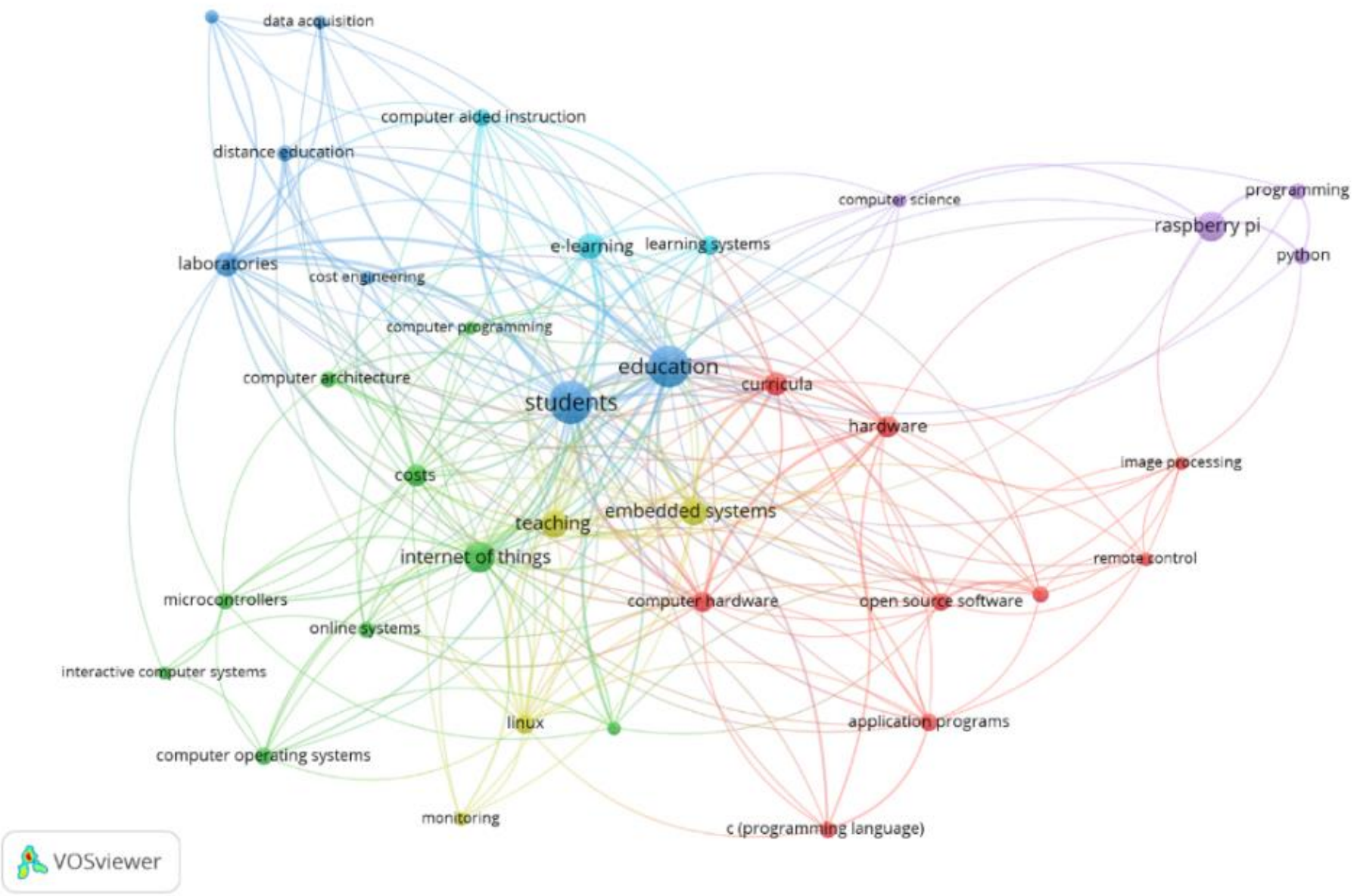

**Figure 3**. Map of clusters for the studies of the synthesis group (*n=154*).

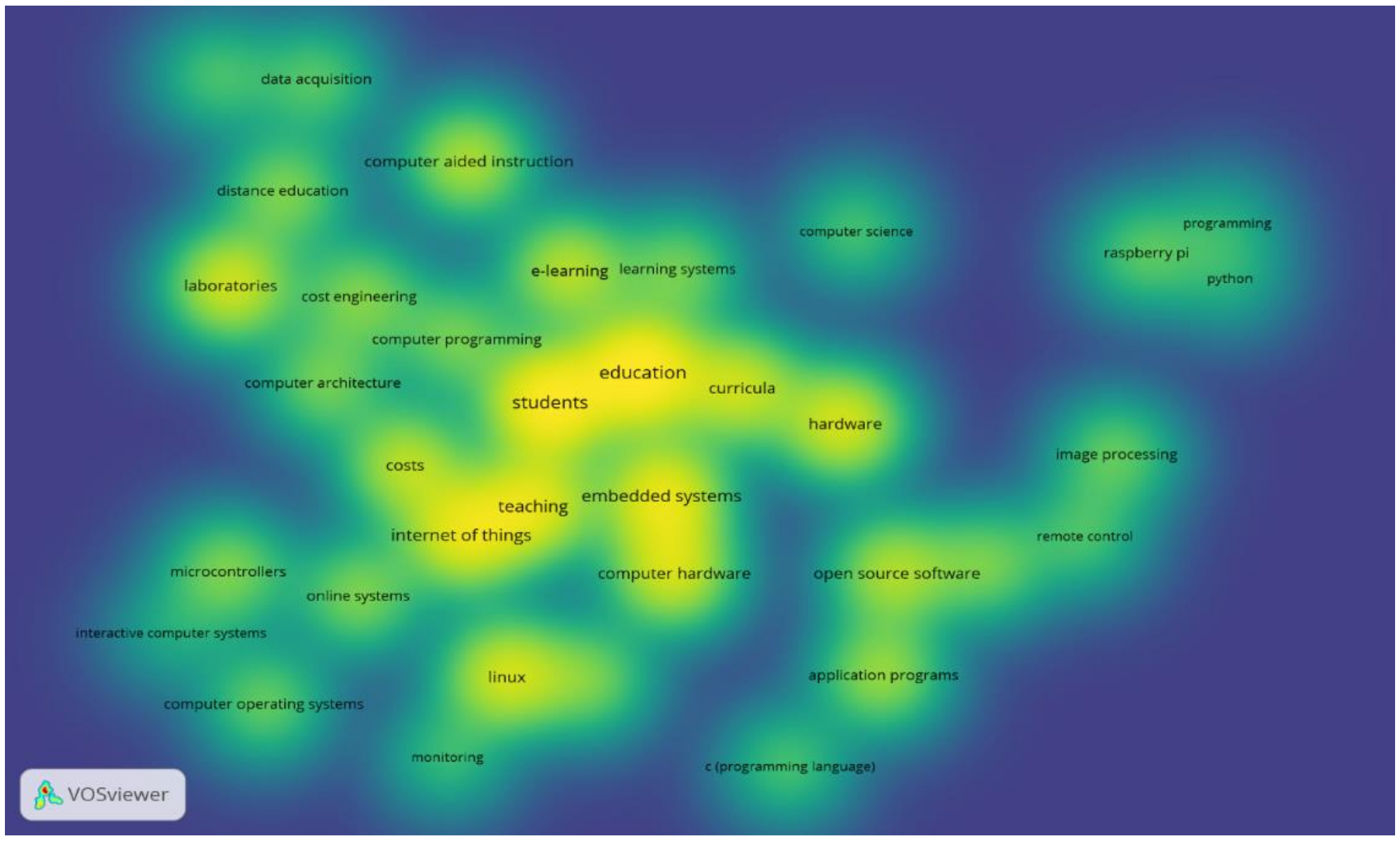

**Figure 4**. Density map for the clusters in figure 3.

**Table 3.** Synthesis of studies by areas and subareas detected in the cluster analysis.

| *Area* | *Subarea* | *References* | *No of studies* |
|---|---|---|---|
| **Laboratories and e-learning** | Control systems and automation | [47–51] | 5 |
| | Robotics and Mechatronics | [52] | 1 |
| | Smart grids | [53,54] | 2 |
| | Improvements and services for remote and virtual laboratories | [55,56] | 2 |
| | Communication networks | [57] | 1 |
| | Bioinformatics | [58] | 1 |
| | Teaching Programming with robotics | [59–61] | 3 |
| **Computing Education** | Parallel computing | [62–65] | 4 |
| | Image Processing | [66] | 1 |
| | Digital literacy | [67] | 1 |
| | Programming in MySQL and assembly | [68,69] | 2 |
| **Robotics** | Teaching design and construction of robots | [70–72] | 3 |
| | Low-cost robotics | [73–75] | 3 |
| | Bioinspired robotics | [76] | 1 |
| | Artificial Intelligence (AI) | [77–79] | 3 |
| **Internet of Things (IoT)** | - | [80–86] | 7 |
| **Persons with disabilities** | - | [87–92] | 6 |

A brief description of the areas is illustrated as follows:

- *Laboratories and e-learning*: Laboratories represent the basis of the experimentation in engineering and computer science and they have formed an active part of the curricula in these disciplines. Students experiment and construct their meanings with hands-on activities, which are the essence of science learning [93]. During the last two decades, the concept of laboratory has been improved by the research about simulated and remote laboratories that currently are used in academic institutions in formal and informal spaces of learning [94]. The costs regarding the physical laboratories and their maintenance, the increasing of students in the classrooms, and the need for a flexible curriculum have yielded to the proliferation of these kinds of laboratories in engineering education [95]. In addition, some detected initiatives in this category have the purpose that the students can access to low-cost and portable laboratories in areas such as digital signal processing, control systems, and smart grids.

- *Computing Education*: Computing Education (CE) is one of the promising fields on SBC in engineering and Computer Science. Educators in this area indicate that SBCs have changed the way to teach complex topics, e.g., parallel computing or bioinformatics. The main part of the proposals in this category count with evidence of application and evaluation. Also, 56.25% of the proposals selected in the group of qualitative synthesis according to Figure 1 belong to this area. Proposals are distributed in the subareas of bioinformatics, parallel computing, robotics and programming, image processing, teaching digital literacy, and teaching MySQL and assembly language.

- *Robotics*: Robotics in higher education represents a good method to encourage learners to study engineering or CS [96]. Moreover, robotics encompasses a multidisciplinary approach in which students can design and learn programming with PhyC. This approach is useful in contextualized computing where educators employ applications or multidisciplinary areas to teach topics in computer science [97]. Features such as a low-cost, multidisciplinary approach, simplified programming environments, and easiness of robot construction have fostered the usage of robotics in higher education [97]. Robotics can help to the innovation in the curricula of engineering and CS with a holistic approach due to its possibility to be included with other thematic areas [98]. Proposals in this area are associated mainly with low-cost robots, bioinspired robotics, and AI.

- *Internet of Things (IoT)*: During the last years, IoT has gained importance in engineering and CS education by two factors: its ubiquitous nature and the requirements of the industry sector in so far as the technical abilities of the students. Madakam et al. [99, p. 165] define IoT as “An open and comprehensive network of intelligent objects that have the capacity to auto-organize, share information, data and resources, reacting and acting in face of situations and changes in the environment”. Although this paradigm is maturing, there exists a lack of educational methodologies in this field. As a result, students in the STEM fields will not be adequately prepared to cope the challenges of the labor market demands [100]. Proposals in this area deal with this matter.

- *Persons with disabilities*: Although the main part of the selected studies in the SLR have been specified for persons without special educational requirements, the studies in this area describe technological proposals or methodologies that can help with learning and teaching in the community of persons with physical and cognitive disabilities.

*4.3 RQ3. What are the main features of the SBCs that support learning and teaching in engineering and computer science education?*

In function of the content analysis, the authors mainly indicate that the incorporation of the SBCs into the curricula of engineering and CS allowed that complex topics, e.g., in parallel computing, cluster design, or image processing that require an expensive infrastructure will be more accessible to the students. This accessibility is provided by the technical features of the SBCs such as low-cost, ease of maintenance, support for multiple programming languages, and the online resources to experiment and learn. Another reported feature of the SBCs is the possibility to experiment and make PhyC by the students from scratch. For many students this feature result important in their learning process, offering the opportunity to experiment in a self-paced way in spaces outside the university, e.g., in their homes. Given the large number of features reported by educators and students, Figure 5 summarizes the most relevant which are classified in the educational and technical levels.

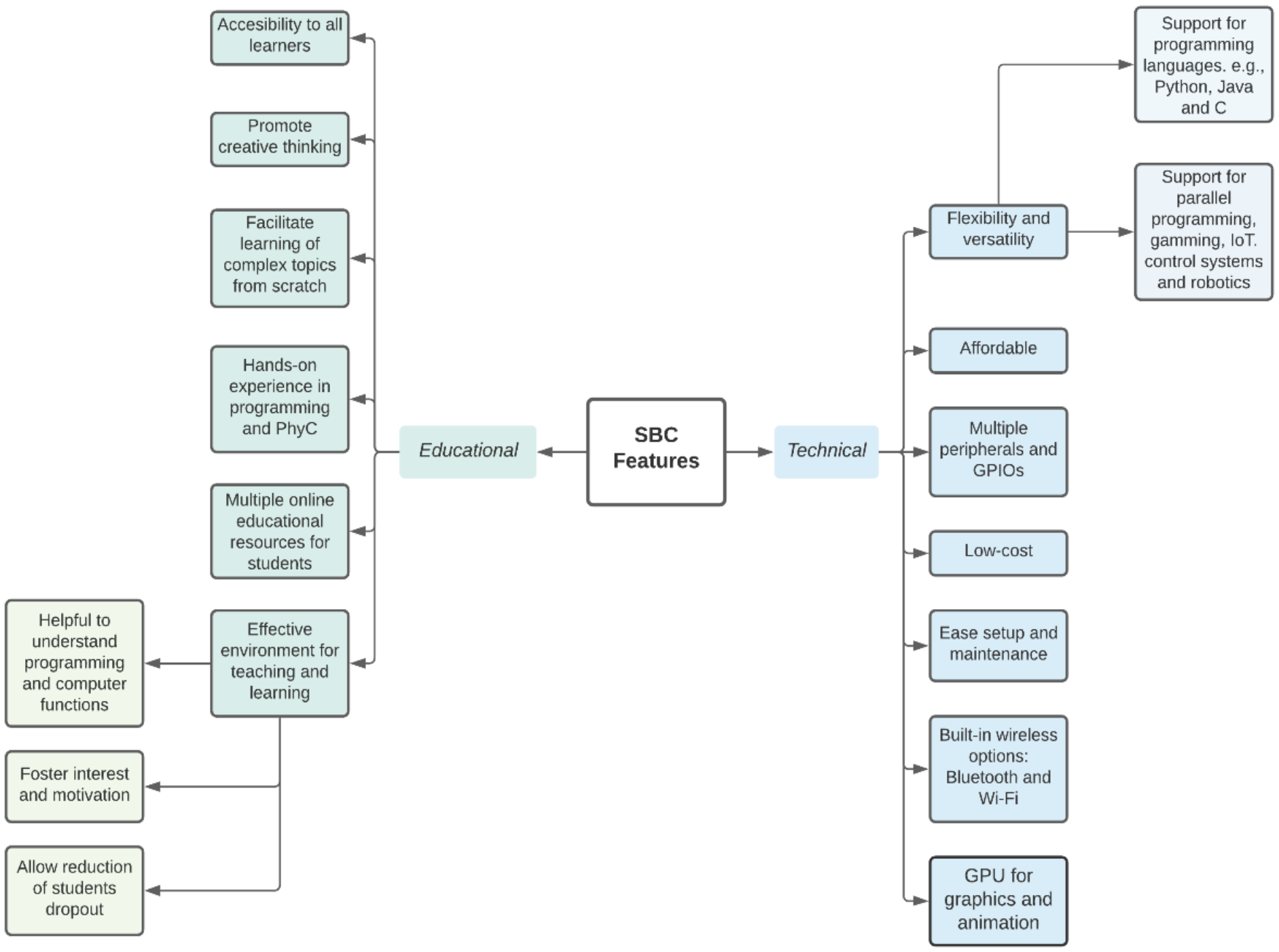

**Figure 5**. Technical and educational SBC features described by the authors.

While these features are positive aspects of the SBCs, some students reported that the problems of hardware and drivers' compatibility delayed the experimentation. This is a factor to considering in the educational design of courses that could affect the learning and motivation of the students.

*4.4 RQ4. What are the learning outcomes and main findings reported by the authors which are supported by the usage of SBCs?*

Table 6 synthesizes the main aspects of the selected proposals in the qualitative synthesis. This table depicts the research approach, educational methodology, participants, analysis techniques, instruments, study aim, and learning outcomes indicated by the authors. Gathering these results, in this section, we address and explain the findings of the authors from the categories of researchers and students' perceptions, educational methodologies, and learning outcomes.

### *4.4.1 Researchers and student perceptions*

As described, educators have felt the need to incorporate SBCs into the curriculum for several reported factors. The first one is the transformation of the curricula mainly in CS regarding programming, parallel computing, cluster development and distributed systems. So, for example, some authors indicate that the curriculum in CS rarely covered topics in parallel computing because of the costs relating to the infrastructure employed. In this situation, students only could access to these devices in specialized laboratories in their universities. With the innovations and features of the SBCs depicted in Figure 5, specially concerning GPU and multicore processors, students can create software to learn the required concepts in an efficient way. Also, students can be exposed to real problems that allow to understand the impact of computing in the society. The second one reported by the authors is the possibility to make "*more experiential*" the learning process of the students with the SBCs. Authors described that not only the lectures and assignments are enough to guarantee learning and grasping of the concepts and topics covered in a course but also the experimentation and interaction. With the experimentation, students can gain skills in the topics or areas that will be required by the labor market even from scratch. Besides, an effect of the experimentation is the increasing of the motivation and the reduction of dropout rates specially for novice programmers and minorities at the undergraduate level. These results are alignment with the implications of the PhyC reported by [20,33]. The third one is relating to the transformation of the teaching methods. The authors claim that while the technical equipment is in continue evolving, the teaching methods remain the same. The lectures and assignments are the typical methods conceived by several teachers in CS or engineering. Thereby, educators have promoted the transformation of the curricula to engage and motivate the students. Through the proposals, the authors encourage to other teachers to take similar pedagogical practices in their courses. The last factor is pertaining to the low-cost and accessibility of the SBCs which is an important feature specially for developing countries. In some studies, [13,101], the SBCs are used to develop solutions to provide efficient laboratories to the students. Authors indicate the problems with low-income and project funding that influence in the employment of the SBCs in education.

Students highlight the interaction with the SBCs as an important factor in their learning process. The hands-on projects, and the peer-to-peer interaction have raised their interest and motivation in the courses. Some students in an intermediate level, for instance, in parallel computing manifest that the SBCs and the interaction with hardware is a powerful tool to increase their motivation. In the same line, the class demonstrations, workshops, and laboratories allow to expand the scope of the computing functions that the students can perform in their courses, stimulating their creativity and curiosity. Hardware malfunctioning, driver´s compatibility, excessive tasks and assignments, and reduced time to interact with the SBCs in class are aspects to improve reported by the students in some proposals. Some students commented the previous aspects as follows:

***S1.*** *The course was very interesting. Being able to utilize a Raspberry Pi was very helpful in the basic understanding of the computer functions.*

***S2.*** *I loved everything about this course. The Raspberry Pi was awesome. Learning Python, scratch all of it was a learning experience, and I loved it.*

***S3.*** *I have experience with parallel processing/programming, but not so much Pis. That's why I took the workshop. I love using the Pis! Wonderfully motivating! Gets students closer to the hardware and powerful enough to motivate studying parallelism. Great workshop!*

***S4.*** *It was a lot harder that it could have been. Didn't first session learnt Python and learnt how to operate outputs. Took time to work, but discovered a driver was missing. Once drivers were in place, it was relatively easy to install requirements.*

***S5.*** *The ARM is a current technology and the Raspberry Pi is super novel, but painful due to hic-ups in the system method for communication/work/with on the Raspberry Pi was over complicated by J-Link.*

Finally, in this subsection, Figure 6 illustrates the word analysis for the perceptions of researchers and students.

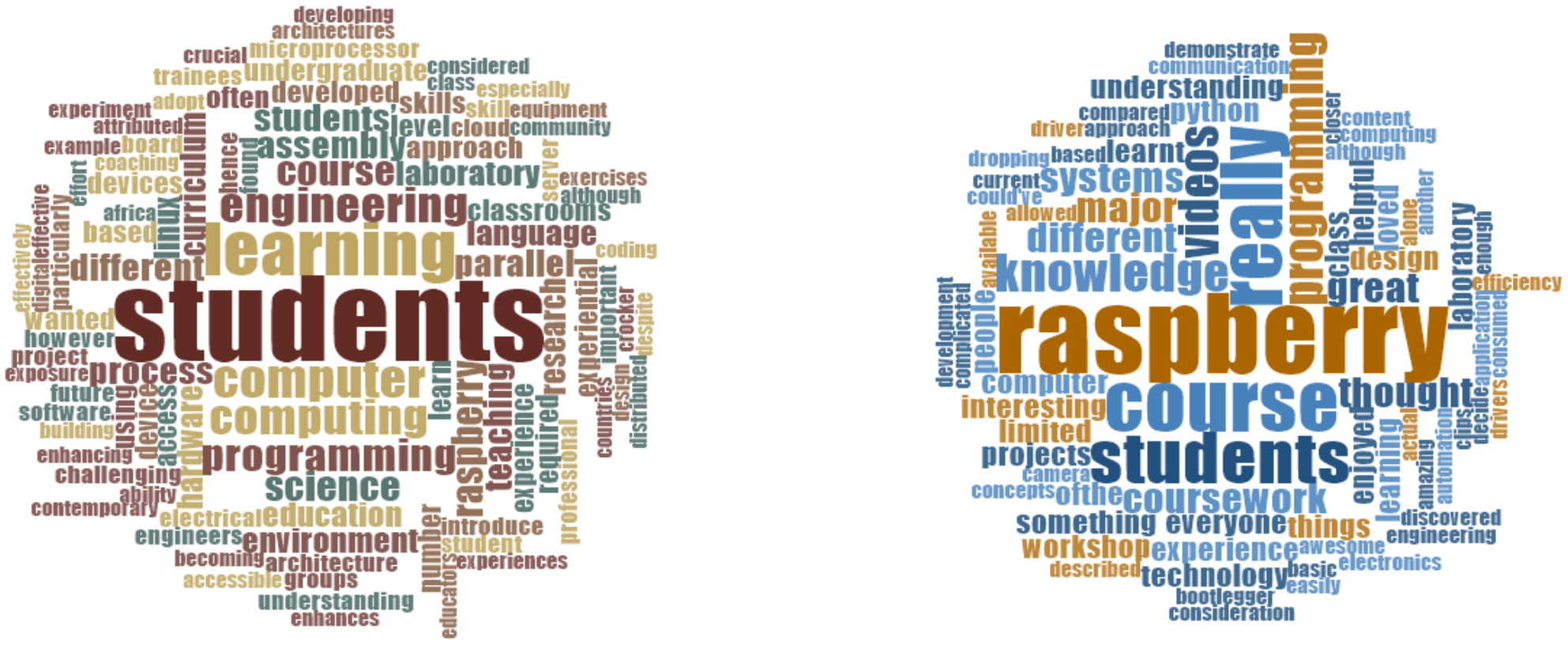

**Figure 6**. Clouds of word frequencies for researchers' perceptions (left) and students' perceptions (right).

*4.4.2 Educational methodologies*

Researchers indicated in their proposals the utilization of the following methodologies:

- Problem-Based Learning (PBL): PBL is an instructional method in which students work in small groups to solve meaningful problems. The teacher is a facilitator to guide the learning process of the students [102]. Four proposals (25%) employ this method.
- Project-Based Learning (PjBL): PjBL is an instructional approach to engage students in investigation. PjBL emphasizes cooperation among students with projects that could require weeks or months [103]. Seven proposals (43.75%) utilize this approach.
- Collaborative Learning (CL): CL is an educational approach to teaching and learning that involves groups of learning that work together to achieve a goal. This goal can be a solution to a problem or product creation [104]. One proposal (6.25%) uses this method.
- Blended Learning: It is a method of educating at distance that combines technology, e.g., the internet with the face-to-face method to teach [105]. One proposal (6.25%) applies this method.
- Flipped Learning: It is a pedagogical teaching approach in which the conventional notion of classroom-based learning is inverted. The students are encouraged to learn the material before class. During the class, the lectures are discussed, and the teachers promote problem-solving [106]. One proposal (6.25%) applies this method.

Furthermore, two proposals (12.5%) not indicated an educational methodology. Previous educational methods are consistent with the constructivism and constructionism theories that deal with the tenet that knowledge and learning are created actively by the students [27,28] through the confluence of the labor of educators and the interaction with the SBCs, in the case of this review.

*4.4.3 Learning Outcomes*

The authors describe learning outcomes in the cognitive and affective domains. Regarding the cognitive learning domain, the authors account for the enhancement of the technical skills of the students in most of the proposals. Increasing of knowledge, comprehension of complex topics, the furtherance of learning, improvement of student’s grades, and integration of knowledge into solving real-world problems are typical cognitive results in the proposals. In the affective learning domain, the improvement of the motivation, course enjoyment and the soft-skills (teamwork, communication, idea generation, evaluation, community engagement, and decision making) required by the students are commonly reported outcomes. Although most of the proposals (87.5%) are situated at the undergraduate level, the study addressed by Wilkinson [107] shows results from the perspective of graduate students who are preparing to be teachers. The proposal highlights the need to promote not only the STEM skills required in the curricula of CS for schools but also the communication and teamwork required by the teachers in the formation process. Also, Ferreira and Freitas [77] describe a robotics course offered to undergraduate, Master and PhD students to enhance learning in kinematics and in Robotic operating System (ROS). Pertaining to the gender and minorities gap in CS, Hallak et al. [108] depict a methodology to engage students that could serve as a reference for future studies.

To collect the information, the authors prefer the surveys in the Likert Scale and semi-structured interviews. The analyses in the studies are performed mainly from a quantitative approach (13 proposals) with descriptive statistics, and 5 studies contrast hypotheses and effect size, employing t-test, ANOVA, Cohen's d, and Pearson's correlation. With respect to these last inferential analyses, see the studies [109–113]. Two studies have a qualitative approach and one a mixed approach that combines learning analytics and semi-structured interviews to collect the information of the students. Participants in the proposals oscillate between 13 to 1214 students with a predominance of the (87.5%) at the undergraduate level and two studies (12.5%) distributed between Master and PhD degrees.

## 5. Limitations of this study

While this SLR had an exploratory nature and it utilized rigorous protocols to collect and analyze the information, there are limitations about the review process and in its scope. Firstly, the search was limited to the databases Ei Compendex, Inspec, and ERIC, using the described searching strings and the described criteria for inclusion and exclusion. However, some SBCs, gray literature, and several peer-reviewed articles in Spanish or Chinese languages were excluded from the review. Likewise, proposals outside the timeframe or published in 2020 later of the final searching process were excluded. However, the main corpus of proposals is in the timeframe of the review. Secondly, selection bias was reduced but not eliminated. This is a proper problem of the SLRs that depends on the agreement and analysis of the coders [14]. Even so, the processes of identification, screening, and eligibility were done jointly with the researchers, discussing the pertinence of each proposal to reduce the selection bias. In stage 2, although 37 papers were shortlisted as case studies, only 16 papers passed the filter of eligibility for the qualitative synthesis. This issue does not reduce the scope of the review, instead evidences a gap of studies in engineering and CS. Thirdly, the areas of application of SBCs in education were generated from the software VOSViewer. In this procedure, we took the most relevant, but others were removed from the process. For instance, some proposals respecting minorities confidence in introductory programming courses, or STEM education can require further studies to include them. Fourthly, the citations of articles and events were retrieved through the Crossref REST API which includes the information of all journal articles, proceedings, or books that are peer-reviewed. Nonetheless, the number of citations can differ from other databases for example SCOPUS or Google Scholar. This discrepancy is due to the type of analysis that performed the API to collect the number of citations that rely on the metadata in each document. Still, the data depicted in the review are consistent and reliable and they can be compared at the Crossref metadata web-site (https://search.crossref.org/). Fifthly, the educational outcomes were classified in cognitive and affective learning domains. These domains cluster the totality of the results reported by the authors. In some systematic reviews, also it is included the behavioral learning domain, but in this review, we have selected the learning domains according to [43].

## 6. Conclusions

In this study, we tried to address the guiding question about how the SBCs are employed in higher education and what learning outcomes are derived from their usage. So, in the review, our interest was twofold. On one hand, we identified the distribution of tertiary levels of the proposals, the educational areas in which SBCs are used, and the learning outcomes in the cognitive and affective domains achieved through their usage. Likewise, we highlighted the opinions of researchers and students as an active part of the review. Since their voices could indicate if the educational methodologies accompanied by the SBCs are relevant for teaching and learning. With this information, we hope that educators and stakeholders can take actions to create educational methodologies and curricula in engineering and CS.

On the other hand, we seek to expand the scope of application concerning the SBCs which are often considered tools for hobbyists and technological amateurs, instead of proper devices for learning and teaching. This aim is highly dependent on the purposes that educators pursue and the notion that they have about the computer and its role in education. Our standpoint as Engineers and Educators is to consider the computer, as any technology in education, just a *tool* [114,115] that should be integrated and reoriented depending on our pedagogical purposes and not the other way around, even though, in many cases this criticism can be overlooked. This fact is supported, e.g., with the low number of proposals that were adequately tackled from an educational approach in this review. A truthful integration and mediation of computers in education only can be attained with the reflection about our educational practices and how computers influence them. Indeed, in this notion from a Vygotskian constructivism perspective, we consider the computers as tools that mediate thought inside a cultural context, operating as elements to promote learning and social interactions as long as it will be understood their purposes and limitations inside the educational field.

Similarly, there are some points that need to be reviewed and investigated. Firstly, it is needed to strengthen the evaluation of proposals from the quantitative, qualitative, or mixed approaches. As described, only 10.4% of the proposals were evaluated adequately. Most of the studies were Technological Proposals (TPs) in the scope of engineering and CS, that is, educational technologies built with SBCs and analyzed under the technical lens. Secondly, some areas such as AI, the impact of programming on the skills of minorities, and STEM education should be examined to broaden the findings of this work. Likewise, the areas of people with disabilities and robotics should be investigated more extensively due to their small number of educational studies. We think that these areas have the potential to be explored in future studies. Thirdly, to find other perspectives different from CS education, it is necessary to create educational methodologies adequately evaluated from other fields in engineering such as robotics, control systems, IoT, or image processing. With the view of these complementary areas, the findings of this SLR can be interpreted from a holistic perspective, helping researchers better understand the role and implications of the SBCs in higher education. Finally, given the current educational context under the COVID-19 pandemic, we believe that the studies in the areas of laboratories and e-learning, and persons with disabilities can help to promote alternatives that benefit students with problems in the access to education and socioeconomic difficulties.

**Appendix section**

- Dataset and supplementary materials for the SLR can be found online at https://github.com/Uniminutoarduino/SBCReview

**Table 4.** Top 20 of most cited articles classified by type of study. Technological Proposal (TP), Lessons Learned (LL), Case Study (CS).

| Authors | Year | Article Title | Citations | Type of study |
|---|---|---|---|---|
| Cox et al. | 2014 | Iridis-pi: a low-cost, compact demonstration cluster | 50 | TP |
| Paull et al. | 2017 | Duckietown: An open, inexpensive and flexible platform for autonomy education and research | 35 | TP |
| He et al. | 2016 | Integrating Internet of Things (IoT) into STEM undergraduate education: Case study of a modern technology infused courseware for embedded system course | 29 | CS |
| Jamieson & Herdtner | 2015 | More missing the Boat; Arduino, Raspberry Pi, and small prototyping boards and engineering education needs them | 27 | LL |
| Ali et al. | 2013 | Technical development and socioeconomic implications of the Raspberry Pi as a learning tool in developing countries | 21 | TP |
| Bermúdez-Ortega et al. | 2015 | Remote Web-based Control Laboratory for Mobile Devices based on EJsS, Raspberry Pi and Node.js | 19 | TP |
| Yamanoor & Yamanoor | 2017 | High quality, low cost education with the Raspberry Pi | 15 | CS |
| Ramirez-Garibay et al. | 2014 | MyVox-Device for the communication between people: blind, deaf, deaf-blind and unimpaired | 12 | TP |
| Krauss | 2016 | Combining Raspberry Pi and Arduino to form a low-cost, real-time autonomous vehicle platform | 11 | CS |
| Peixoto et al. | 2018 | Robotics tips and tricks for inclusion and integration of students | 11 | LL |
| Reck & Sreenivas | 2015 | Developing a new affordable DC motor laboratory kit for an existing undergraduate controls course | 11 | TP |
| Barker et al. | 2013 | 4273π: Bioinformatics education on low cost ARM hardware | 10 | CS |
| Klinger & Madritsch | 2015 | Collaborative learning using pocket labs | 10 | CS |
| Chaczko & Braun | 2017 | Learning data engineering: Creating IoT apps using the node-RED and the RPI technologies | 8 | TP |
| Kawash et al. | 2016 | Undergraduate Assembly Language Instruction Sweetened with the Raspberry Pi | 8 | CS |
| He et al. | 2017 | Internet of Things (IoT)-based Learning Framework to Facilitate STEM Undergraduate Education | 7 | CS |
| Luis Guzmán et al. | 2015 | Teaching Control Engineering Concepts using Open Source tools on a Raspberry Pi board | 7 | TP |
| Marot & Bourennane | 2017 | Raspberry Pi for image processing education | 7 | TP |
| Vasilchenko et al. | 2017 | Media Literacy as a By-Product of Collaborative Video Production by CS Students | 7 | CS |
| Garcia et al. | 2019 | An Approach of Training Virtual Environment for Teaching Electro-Pneumatic Systems | 6 | TP |

**Table 5.** Main software utilized by the researchers.

| Software | Description | Web page |
|---|---|---|
| Node.js | Open-source, cross-platform, JavaScript runtime environment that executes JavaScript code outside a web browser. | https://nodejs.org/en/ |
| Easy JavaScript Simulations (EJsS) | Free authoring tool written in Java that helps non-programmers create interactive simulations in Java or JavaScript, mainly for teaching or learning purposes. | https://www.um.es/fem/EjsWiki/ |
| WebIOPi | Python package to access remotely to GPIO with support to protocols such as Serial, SPI and I2C. | http://webiopi.trouch.com/ |
| REX Control System | Environment to create real-time algorithms for control systems employing Raspberry Pis. | https://www.rexcontrols.com/articles/getting-started-with-rex-on-raspberry-pi |
| Mozaïk | Declarative front-end JavaScript library for building user interfaces. | http://mozaik.rocks/ |
| Motion | Highly configurable program that monitors video signals from many types of cameras. It supports network and Pi cameras through RTSP, RTMP, and HTTP protocols. | https://motion-project.github.io/motion_config.html |
| High-Performance Portable MPI (MPICH2) | High performance and widely portable implementation of the Message Passing Interface (MPI) standard. | http://www.mpich.org/ |
| LeJos | Operating system for LEGO robots. | http://www.lejos.org/ |
| OpenMP | Implementation of multithreading, a method of parallelizing whereby a master thread (a series of instructions executed consecutively) forks a specified number of slave threads and the system divides a task among them. | https://www.openmp.org/ |
| BLAST | It compares nucleotide or protein sequences to sequence databases and calculates the statistical significance. | https://blast.ncbi.nlm.nih.gov/Blast.cgi |
| Xinu | Small, elegant operating system that supports dynamic process creation, dynamic memory allocation, network communication, local and remote file systems, a shell, and device independent I/O functions. | https://xinu.cs.purdue.edu/ |
| Slack | Package for messaging in working teams. | https://slack.com/intl/es-co/features |
| OpenCV | Highly optimized library with focus on real-time applications. | https://opencv.org/ |
| Bootlegger | It bridges the gap between professional filmmakers and people with no prior filming experience wishing to record video on their mobile phone. | https://openlab.ncl.ac.uk/research/bootlegger-citizen-filmmaking-app/ |
| Stage Robot Simulator | It provides a virtual world populated by mobile robots with various objects to sense and manipulate. | http://playerstage.sourceforge.net/doc/Stage-3.2.1/ |
| Gazebo | It offers the ability to simulate populations of robots accurately and efficiently in complex indoor and outdoor environments. | http://gazebosim.org/ |
| MORSE | Generic simulator for academic robotics. It focuses on realistic 3D simulation of small to large environments, indoor or outdoor, with one to tenths of autonomous robots. | https://www.openrobots.org/morse |
| Node-RED | Programming tool for wiring together hardware devices, APIs and online services in new and interesting ways. | https://nodered.org/ |
| Robotic Operating System (ROS) | Real-time operating system for robotics. | https://www.ros.org/ |

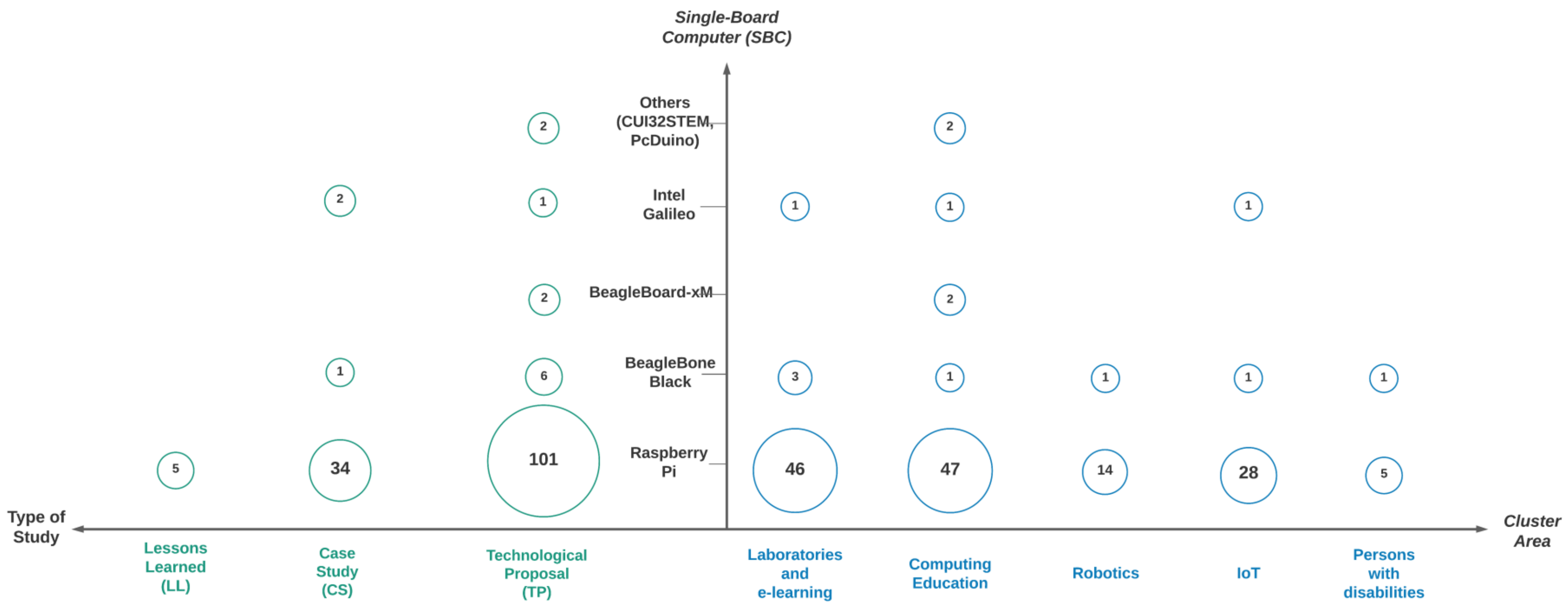

**Figure 7**. Scatter plot for the cluster areas vs. SBCs, and type of study vs. SBCs. (*n=154*).

**Table 6.** List of selected studies for qualitative synthesis (*n=16*).

| Title | Reference | Educational methodology | Research approach and instruments | Participants, tertiary level | Study aim | Learning outcomes (cognitive and affective domains) |
|---|---|---|---|---|---|---|
| *Integrating Hardware Prototyping Platforms into the Classroom* | [75] | Project Based-Learning | Quantitative, survey | 53 students, (undergraduate) | Investigate the usefulness of integrating low cost open-source hardware platforms into engineering and computer science courses. | *Cognitive*: Improvement of students' grades in comparison with the non-participant group. students' learning curve is constantly increasing throughout the semester. *Affective*: Improvement of student motivation. |
| *Teaching on Demand: an HPC Experience* | [116] | Project-Based Learning | Qualitative, survey | 18 students, (undergraduate) | Increase knowledge about High-Performance Computing (HPC) in the students of engineering and computer science. | *Cognitive*: HPC knowledge increased among students. Students recognized that there are other uses of HPC different to merely complex computations. *Affective*: Enhancement of interest and motivation towards HPC. |
| *Incorporating a Raspberry Pi into a Computer Information Systems Initial Course* | [117] | Problem-Based learning | Quantitative, survey | 1214 students, (undergraduate) | Incorporate Raspberry Pi boards as part of a curriculum renovation of the course COMP100 offered to students of the College of Engineering and Information Sciences. | *Cognitive*: Improvement of the students' grades with an increase of (6%) in the pass rate in comparison with the first COMP100 course. *Affective*: Students satisfaction with the new course. |
| *Robotics as multidisciplinary learning: a summer course perspective* | [77] | Collaborative learning | Quantitative, survey | 84 students, (undergraduate, MSc, PhD) | Introduce students to robotics in a summer course of two months. The topics addressed were programming, sensors, artificial intelligence (IA) and robot design. | *Cognitive*: Increasing of knowledge and understanding in the topics concerning Robotic Operating System (ROS) and Kinematics. |
| *The Impact of Incorporating Hands-on Raspberry Pi Projects with Undergraduate Education in Boosting Students' Interest* | [108] | Project-Based Learning | Quantitative, survey | 37 students, (undergraduate) | Study the impact of hands-on activities in the curriculum of CS, making it more attractive to students from minorities. | *Cognitive*: Improvement of the students' grades. *Affective*: Reduction of fear regarding computer science major, increasing of motivation to problem-solving. |
| *An Energy-Efficient Secure Adaptive Cloud-of-Things (CoT) Framework to Facilitate Undergraduate STEM Education* | [81] | Project-Based Learning | Quantitative, survey | 13 students, (undergraduate) | Transform the STEM curricula, making that the students understand the IoT/CoT concepts and gain hands-on programming experience. | *Cognitive*: Students integrated knowledge of CoT and IoT into solving real-world problems. |

**Table 7.** Table 6 continued.

| Title | Reference | Educational methodology | Research approach and instruments | Participants, tertiary level | Study aim | Learning outcomes (cognitive and affective domains) |
|---|---|---|---|---|---|---|
| *CS0: Introducing computing with Raspberry Pis* | [118] | Project-Based Learning | Quantitative, survey | Students of six majors related to software engineering and information systems (Number not indicated), (undergraduate) | Redesign a course focused on the use of Raspberry Pi as a common computing platform to encourage student experimentation. | *Cognitive*: Students understood concepts in computing including programming, networking, data representation. *Affective*: Students were active participants in the computer science community. |
| *Undergraduate Assembly Language Instruction Sweetened with the Raspberry Pi* | [109] | Problem-Based Learning | Quantitative, survey | 337 students, (undergraduate) | Reducing the frustration and demotivation that students experience when dealing with assembly language in the course of computer machinery II. | *Cognitive*: Improvement of students' grades, students consider that the Raspberry Pi (RPi) enhance their learning in the course. *Affective*: Students enjoyed working with RPi and they were motivated to learn through its usage. |
| *Portable Parallel Computing with the Raspberry Pi* | [110] | Not indicated | Quantitative, survey | 55 students, (undergraduate) | Discuss the usage of the Raspberry Pi single-board to provide hands-on learning experiences in parallel computing. | *Cognitive*: Enhancement of learning in parallel computing in topics such as decomposition, algorithms, and architecture. *Affective*: Motivation and enjoyment to learn parallel computing using the Raspberry Pi. |
| *Teaching of IA-32 Assembly Language Programming Using Intel Galileo* | [111] | Project-Based Learning | Quantitative, survey | 44 students, (undergraduate) | Transform the traditional course of assembly using processors Intel 8086 and the SBC (Intel Galileo) in order to motivate the students to learn assembly language for microprocessors. | *Affective*: Students were satisfied with the usage of Intel Galileo to complete their assignments and laboratories. |
| *Enhancing Students' Learning Experience via Low-Cost Network Laboratories* | [112] | Problem-Based Learning | Quantitative, survey | 40 students, (undergraduate) | Design and development of a low-cost laboratory to improve students' learning experience in distributed computing systems. | *Cognitive*: Enhancement of the students' learning with the usage of the laboratory. Impact in knowledge and in previous experience in distributed systems. |

**Table 8.** Table 6 continued.

| Title | Reference | Educational methodology | Research approach and instruments | Participants, tertiary level | Study aim | Learning outcomes (cognitive and affective domains) |
|---|---|---|---|---|---|---|
| *Media Literacy as a By-Product of Collaborative Video Production by CS Students* | [119] | Flipped Learning | Mixed, learning analytics, semi-structured interview | 34 students, (undergraduate) | Investigate how undergraduate students of computer science can learn media literacy as a by-product of collaborative video tutorial production. | *Cognitive*: Students understood the reusability of media components such as clip authorship, visual and audio quality. *Affective*: Students collaborated and worked as a team. |
| *A low-Cost Laboratory for Enhanced Electrical Engineering Education* | [101] | Not indicated | Quantitative, survey | 81 students, (undergraduate) | Develop a low-cost laboratory to enhance the teaching of electrical and electronic engineering. | *Cognitive*: Improvement of Learning in the digital signal processing concepts of (Sampling, Discrete Fourier Transform, autocorrelation, fundamental frequency estimation). *Affective*: Laboratory exercises motivated the students to learn the course material, students enjoyed the laboratories. |
| *A Triumvirate Blended Learning Method for Embedded Computational Devices used in the Internet of Things: A Case Study* | [120] | Blended-Learning | Quantitative, survey | 117 students, (undergraduate) | Propose a new hybrid-learning model to instruct mid-level students in electrical and information engineering. | *Cognitive*: Students perceived an increase of their programming ability. *Affective*: Increased motivation and enthusiasm to learn Engineering. |
| *Implementing a Cross-Curricular Digital Project into a PGCE Computer Science Initial Teacher Education Course* | [107] | Problem-Based Learning | Qualitative, semi-structured interview | 76 graduate students, (Initial teacher trainees.) | Open the discussion over the processes and skills of the subject of Computer Science in the new English curriculum of 2013. | *Affective*: Enhancement of the soft skills for initial teacher trainees. |
| *Case Study: Using Project Based Learning to Develop Parallel Programming and Soft Skills* | [113] | Project-Based Learning | Quantitative, survey | 124 students, (Undergraduate) | Investigate the effectiveness of using Problem-Based Learning to teach parallel programming. | *Cognitive*: Improvement in the parallel programming skills of the students. *Affective*: Students improved their soft skills (teamwork, communication, evaluation and decision-making). |